\documentstyle[12pt]{article}
\makeatletter

\@addtoreset{equation}{section}
\makeatother
\def\abstract#1{{\centerline{\bg Abstract}} \vskip 3mm \par #1}
\def\cy{Calabi-Yau}
\def\cym{Calabi-Yau manifold}
\def\lg{Landau-Ginzburg}

\def\p{Poincar\'{e}}

\def\tg{\tilde{\theta_{i}}^{g}}

\def\thd{\tilde{\theta_{i}}^{h^{\prime}}}

\def\inbar{\vrule height1.5ex width.4pt depth0pt} 

\def\ZZ{\relax{\sf Z\kern-.4em \sf Z}}  \def\IR{\relax{\rm  
I\kern-.18em R}}
\def\IN{\relax{\rm I\kern-.18em N}} \def\IP{\relax{\rm I\kern-.18em  
P}}
\def\IQ{\relax\,\hbox{$\inbar\kern-.3em{\rm Q}$}}
\def\la{\lambda}
\def\IC{\hbox{\,$\inbar\kern-.3em{\rm C}$}}

\def\({\lbrack}
\def\){\rbrack}

\def\ketc#1{{\left| #1\right\rangle}_{\rm (c,c)}}
\def\keta#1{{\left| #1\right\rangle}_{\rm (a,c)}}

\begin{document}
\baselineskip=6mm
\begin{flushright}
{hep-th/9705012} \\
{OU-HET 263} \\
{April 1997}
\end{flushright}
\vskip 1.5cm
\centerline{\LARGE {\bf Non-Perturbative Superpotentials in }}
\vskip 0.5cm
\centerline{\LARGE {\bf \lg\ Compactification }}
\vskip 1.5cm
\centerline{\large Hitoshi \ Sato}
\vskip 1cm
\centerline{\it Department of Physics, Faculty of Science, Osaka  
University}
\centerline{\it Toyonaka, Osaka 560, Japan} 
\centerline{email address : sato@funpth.phys.sci.osaka-u.ac.jp}
\vskip 1.5cm
\centerline{\large {\bf ABSTRACT} }

\vskip 0.5cm

We study the \lg\ models which correspond to \cy\ four-folds.
We construct the index of the typical states 
which correspond to toric divisors.
This index shows that whether a corresponding divisor can generate 
a non-perturbative superpotential.
For an application, we consider
the phase transition
in terms of the orbifold constraction.
We obtain the simple method by which
the divisor, which can not generate a superpotential
in the original theory,
can generate a superpotential after orbifoldization.

\thispagestyle{empty}

\clearpage

\pagenumbering{arabic}

\section{Introduction}
Recently the \cy\ four-folds
have been paied attention to
in the context of
M-theory \cite{witten1,schwarz}
and F-theory
\cite{mv} compactification.
Witten found \cite{witten}
the interesting property of
the \cy\ four-fold compactification,
i.e. the special class of the divisors with wrapped 5-branes
can generate the $N = 1$ superpotential.
The condition whether or not a divisor ${\cal D}$
can genarete a superpotential
is purely topological,
i.e. the Euler character
$\chi ({\cal D}) = 1$ is necessary 
(not sufficient) and
the holomorphic Hodge numbers
$h^{1,0}({\cal D}) =
h^{2,0}({\cal D})=
h^{3,0}({\cal D}) = 0$
are sufficient
(not necessary) conditions.

In this note,
we study  the such topological properties of divisors
in terms of \lg\ models.
It is well known that 
the \lg\ models as well as their orbifolds
with central charge $c = 3d$
can describe the \cy\ manifolds of dimension $d$.
For example, the Hodge numbers $h^{i,j}$ 
can be identified with
the numbers of the $(c,c)$ or $(a,c)$ states
with $U(1)$ charge 
$(d-i,j)$ or $(-i,j)$ respectively,
where $c$ $(a)$ denotes the (anti-)chiral ring
of $N = 2$ conformal field theory
\cite{lvw}.
These numbers can be calculated
without any geometrical information
and the simple formulas were obtained \cite{sa1}
for three-fold case.
Since four-folds are of our interest,
we will consider the \lg\ models (and their orbifolds)
with $c = 12$.

For geometrical description of the \cym s,
toric geometry is most useful.
In the work of
Klemm, Lian, Roan and Yau \cite{klry},
the topological properties for the divisors
is described in terms of toric geometry.
It is of our interest 
to find the corresponding result
in terms of \lg\ models.
To do this,
we will use the method obtained in \cite{sa3,sa4}
where it was shown how these two description,
the toric geometry and the \lg\ models,
can be identified.
Once we obtain the \lg\ description,
we can find the simple conditions for the states
which correspond to the divisors.
These conditions, as expected,
can be applied straightforwardly to the calculation
without any geometrical information.
Moreover, by the orbifold construction of \cite{sa4},
we can easliy find 
the new model
in which
the divisor, which can not generate the superpotential
in the original theory,
{\it can} generate the superpotential
in the orbifoldized theory.
This implies that the phase transition
in terms of our orbifoldization
can be characterized by the property of the divisors
which lead to
the non-perturbatively generated superpotentials.
Similar transitions are discussed in \cite{bls}.

This papre is organized as follows.
In section 2, we briefly review the results of \cite{klry}
and their \lg\ interpretaion is obtained in section 3.
Moreover in section 3, we develop the simple method
to dicide whether or not 
a divisor with wrapped 5-brane
can generate a superpotential.
In the last section
we consider the \lg\ orbifolds,
which correspond to the \cy\ four-fold orbifolds,
and their phase transitions.
We can easily find the appropriate orbifoldization 
by the method obtained in \cite{sa4},
such that
the divisor which can not generate the superpotential
in the original theory
can generate the superpotential
in the orbifoldized theory.

\section{Toric description of \cy\ four-folds
and their divisors}
The topological numbers for \cy\ four-folds
are studied in \cite{SethiVafaWitten,klry,Mohri}.
At first sight, the non-trivial Hodge numbers are
$h^{1,1}$, $h^{3,1}$, $h^{2,1}$ and $h^{2,2}$
(up tp dualities).
However by the index theorem \cite{SethiVafaWitten}
it is shown that there is one relation
\begin{equation}
h^{2,2} = 2(22 + 2 h^{1,1} + 2 h^{3,1}  -  h^{2,1}),
\end{equation}
so that only three Hodge numbers are independent,
namely $h^{1,1}$, $h^{3,1}$ and $h^{2,1}$.
It will be clear that 
this fact is important
when one considers the relation
between the \lg\ descripton
and the toric geometry.
The Euler number can be simply written as
\begin{equation}
\chi = 6 (8 +  h^{1,1}  +  h^{3,1}  -  h^{2,1} ).
\end{equation}

There are formulas for these Hodge numbers
$h^{1,1}$, $h^{3,1}$ and $h^{2,1}$
in terms of toric geometry.
The formulas of the first two of these
are obtained
due to Batylev's original work \cite{vb}:
\begin{equation}
\label{vb11}
h^{1,1} = 
l(\Delta^{*}) - 6 
- \sum_{\bf{codimension} \; {\Theta^{*}} = 1}
{l^{'}({\Theta^{*}})}
+\sum_{\bf{codimension} \; \Theta^{*} = 2, 
\Theta^{*} \in \Delta^{*}}
{l^{'}(\Theta^{*})
l^{'}(\Theta)}
,
\end{equation}
\begin{equation}
\label{vb31}
h^{3,1} = 
l(\Delta) - 6 
- \sum_{\bf{codimension} \; {\Theta} = 1}
{l^{'}({\Theta})}
+ \sum_{\bf{codimension} \; \Theta = 2, 
\Theta \in \Delta}
{l^{'}(\Theta)
l^{'}(\Theta^{*})}
,
\end{equation}
where 
${l({\Delta})}$ ($l(\Delta^{*})$)
denotes the number of integral points
in the Newton polyhedron ${\Delta}$ 
(the dual polyhedron ${\Delta}^{*}$)
and 
${l^{'}({\Theta})}$ ($l^{'}(\Theta^{*})$)
denotes the number of integral points
interior of the face ${\Theta}$ 
(the dual face ${\Theta}^{*}$).
The formula for $h^{2,1}$ is obtained 
in \cite{klry,Mohri} to be
\begin{equation}
\label{vb21}
h^{2,1} = 
\sum_{\bf{codimension} \; \Theta = 3, 
\Theta \in \Delta}
{l^{'}(\Theta)
l^{'}(\Theta^{*})}.
\end{equation}

The authors of ref. \cite{klry} studied
the topological numbers of the divisors in \cy\ four-folds
using toric geometry.
They analyze the local structure of the divisors
coming from the blowing up of the singularities
on the hypersurface of the \cy\ embedded in 
the weighted complex projective space.
They have classified the divisors
by thier topological numbers.
We will briefly review their results
which we need.
\begin{description}
\item[Case A] ${d_{{\Theta_{k}^{*}}} = 3}$, \ \ 
 $h^{0,0}({\cal D}_{k}) = l^{\prime}(\Theta_{k}) + 1$. \ \
 This case explains the additional fourth therm in (\ref{vb11}).
\item[Case B] ${d_{{\Theta_{k}^{*}}} = 2}$, \ \ 
 $h^{0,0}({\cal D}_{k}) = 1$, \
 $h^{1,0}({\cal D}_{k}) = l^{\prime}(\Theta_{k})$. \ \
 In this case we get 
${l^{\prime}(\Theta_{k}^{*})} \cdot {l^{\prime}(\Theta_{k}^{*})}$
$(3,2)$ forms, their dual $(1,2)$, $(2,3)$ and $(2,1)$ forms 
on a four-fold,
where we have used the \p\ and complex cojugation dualities.
\item[Case C] ${d_{{\Theta_{k}^{*}}} = 1}$, \ \ 
 $h^{0,0}({\cal D}_{k}) = 1$, \
 $h^{2,0}({\cal D}_{k}) = l^{\prime}(\Theta_{k})$. \ \
 On a four-fold, we get additional $(3,1)$ forms 
of the fourth term in (\ref{vb31}).
\end{description}
and other $h^{i,j}({\cal D}_{k}) = 0$
in all the cases.
As a result, they find the simple formula for the Euler number
of the divisor ${\cal D}_{k}$, i.e.
\begin{equation}
\label{chi}
\chi ({\cal D}_{k}) =
1 +  (-1)^{{\rm dim}{{\Theta}_{k}} + 1}
\ l^{\prime} ({{\Theta}_{k}}),
\end{equation}
where $l^{\prime} ({{\Theta}_{k}})$ is the number of integral points

in the face ${{\Theta}_{k}}$ 
which is dual to the point in the dual polyhedron
$\Delta^{*}$ corresponding to
the divisor ${\cal D}_{k}$  of our interest.

\section{\lg\ analysis}
A \lg\ model is characterized by a superpotential
$W(X_{i})$ where $X_{i}$ are $N = 2$ chiral superfields.
The \lg\ orbifolds 
\cite{iv} are obtained by
quotienting with an Abelian symmetry group $G$ of 
$W{(X_{i})}$,
whose element $g$ acts as an $N \times N$ diagonal matrix,
$g: X_{i} \rightarrow e^{2 \pi i {\tg}}X_{i}$,
where $0 \leq \tg < 1$.
Of course the $U(1)$ twist \ 
$j: X_{i} \rightarrow e^{2 \pi i {q_{i}}}X_{i}$ \ 
generates the symmetry group of
$W{(X_{i})}$,
where $q_{i} = {w_{i} \over d}$,\ \ 
 $W(\la^{w_{i}} X_{i}) = \la^{d}W(X_{i})$ \ and
$\la \in \IC^{\ast}$.
Using the results of Intriligator and Vafa \cite{iv},
we can construct the $(c,c)$ and $(a,c)$ rings,
where $c$ ($a$) denotes chiral (anti-ciral).
Also we could have the left and right $U(1)$ charges of 
the ground state \  
$\keta h$ \ 
in the $h$-twisted sector 
of the $(a,c)$ ring.
In terms of spectral flow, \ $\keta h$ is mapped to the (c,c)
state $\ketc {h^{\prime}}$ with $h^{\prime} = hj^{-1}$.
Then the charges of the (a,c) ground state of $h$-twisted sector
$ \keta {h} $
are obtained to be
\begin{equation}
\label{uac}
\begin{array}{cc}
\left(\begin{array}{c}
J_{0} \\
\bar{J_{0}}
\end{array} \right) &
\end{array}
\keta {h} 
= 
\begin{array}{cc}
\left(\begin{array}{c}
{ - \sum_{\thd>0}{(1-q_{i}-\thd)}}
+ \sum_{\thd=0} {(2q_{i}-1)}
 \\
{ \sum_{\thd>0}{(1-q_{i}-\thd)}}
\end{array} \right) &
\keta {h}.
\end{array}
\end{equation}

Our purpose of this section is 
to consider the topological properties
of the divisors in terms of the \lg\ model.
As we reviewed in the previous section,
there are three classes of divisors
of our interest
which are called Case A, B and C.
What is the corresponding classification 
in the \lg\ description?
To answer this question, 
we should recall the observation
in ref.\cite{sa4}
where the \cy\ three-folds are considered 
in terms of the \lg\ model.

The observation of \cite{sa4} is
that if the $(-1,1)$ state
$\keta {h^{\prime}}$ exists
in the ${h^{\prime}}$-twisted sector
then it is possible to exist
the states written in the form
$\prod_{\thd = 0}{X_{i}^{l_{i}}} \keta h$
in the $h$-twisted sector,
where $h^{\prime} \equiv hj^{-1}$.
The $U(1)$ charge of the state
$\prod_{\thd = 0}{X_{i}^{l_{i}}} \keta h$
depends on 
the number of invariant fields
under the $h^{\prime}$ action.
We denote by $I_{h^{\prime}}$
that number of invariant fields.
For three-folds,if $I_{h^{\prime}} = 2, \  3$,
then the possible $U(1)$ charges are
$(-1,1)$, $(-2,1)$, respectively.

Applying this obsevation to our four-fold case,
we obtain the following results
which correspond to 
the previous classification of divisors.
\begin{description}
\item[$I_{h^{\prime}} = 2$] It is possible to exist 
$(-1,1)$ states $X_{i}^{l_{i}}X_{j}^{l_{j}} \keta h$.
This case corresponds to the Case A. 
\item[$I_{h^{\prime}} = 3$] It is possible to exist 
$(-2,1)$ states $X_{i}^{l_{i}}X_{j}^{l_{j}}X_{k}^{l_{k}} \keta h$.
This case corresponds to the Case B. 
\item[$I_{h^{\prime}} = 4$] It is possible to exist 
$(-3,1)$ states 
$X_{i}^{l_{i}}X_{j}^{l_{j}}X_{k}^{l_{k}}X_{m}^{l_{m}} \keta h$.
This case corresponds to the Case C. 
\end{description}
where at least one $l_{i} > 0$.
Off course, we have assumed
the existence of the $(-1,1)$ state
$\keta {h^{\prime}}$.
The well-known identification
between $(i,j)$ forms and $(-i,j)$ states
implies 
that we can classify
by the number $I_{h^{\prime}}$
the $(-1,1)$ states
which correspond to the divisors.
We will study each case more in depth
and find the condition of the $(-1,1)$ states
whether the corresponding divisors
can generate the superpotntial.

First, consider the case of $I_{h^{\prime}} = 2$.
In this case, all the divisors can generate
the superpotentials.
Since it is shown \cite{sa3,sa4} that 
a $(-1,1)$ state $\keta {h^{\prime}}$
corresponds to a integral point 
in the dual polyhedron $\Delta^{*}$,
the number of $(-1,1)$ states 
$X_{i}^{l_{i}}X_{j}^{l_{j}} \keta h$
in the $h$-twisted sector is equal to
$l^{\prime}(\Theta)$.

Let us turn to the $(c,c)$ ring by spectral flow.
For the four-fold case,
the states which correspond to the $(1,1)$ forms 
have thier $U(1)$ charge being $(3,1)$.
Using the technique developed in \cite{sa1},
we obtain the index 
$\beta (h^{\prime})$
which counts the number of $(3,1)$ states
written in the form
$X_{i}^{l_{i}}X_{j}^{l_{j}} {\ketc {h^{\prime}}}$,
i.e.
\begin{equation}
\label{beta}
\beta (h^{\prime}) = 
\frac{1}{| \rm G |}
\sum_{ {\rm all \ g \ \in \ G}}{
\prod_{ \tg = \thd = 0}{
(1 - \frac{1}{q_{i}})}
}.
\end{equation}
(If the set of $i$'s satsfying 
$\tg = \thd = 0$ is empty,
then we define 
$\prod_{ \tg = \thd = 0}{
(1 - \frac{1}{q_{i}})}
 = 1$
).
Clearly, in this case we have 
$\beta (h^{\prime}) =
l^{\prime}({{\Theta}_{k}})$.

The above discussions can be extended to 
other $(-1,1)$ states $\keta {h^{\prime}}$
with 3 or 4 invariant fields under 
$h^{\prime}$ action.
Remember that if there are 3 (4) invariant fields
under $h^{\prime}$ action,
it is possible to exist $(-2,1)$ ($ \ (-3,1) \ $) states
$\prod_{\thd = 0}{X_{i}^{l_{i}}}{\keta h}$.
For the case with 
$I_{h^{\prime}} = 4$,
the index $\beta (h^{\prime})$ is defined
as in (\ref{beta})
over the $(c,c)$ ring 
in the $h^{\prime}$-twisted sector.
However in the case with
$I_{h^{\prime}} = 3$,
the index $\beta(h^{\prime})$
should be defined to be
\begin{equation}
\label{beta2}
\beta (h^{\prime}) = 
\frac{-1}{2}
\frac{1}{| \rm G |}
\sum_{ {\rm all \ g \ \in \ G}}{
\prod_{ \tg = \thd = 0}{
(1 - \frac{1}{q_{i}})}
},
\end{equation}
since in the same twisted sector 
there are the $(-2,1)$ and $(-1,2)$ states
which corresponds to 
the $(2,1)$ and $(1,2)$ forms
obeying the complex conjugation
duality.

Finally we obtain
\begin{equation}
\label{betal}
\beta (h^{\prime}) =
l^{\prime} ({{\Theta}_{k}}),
\end{equation}
in all the cases with
$I_{h^{\prime}} = 2, \ 3, \ 4$.
Note that
${{\rm dim}{{\Theta}_{k}}+ 1} =
I_{h^{\prime}}$.
Comparison of (\ref{chi}) with (\ref{betal})
leads us to conclude that
a non-perturbative superpotential can be generated if
\begin{equation}
\beta (h^{\prime}) = 0.
\end{equation}
This condition is very useful 
because the calculation 
can be done straightforwardly without
any knowledge of geometry.
Moreover, our condition can be available 
for the \lg\ orbifolds which correspond to
the \cy\ four-fold orbifolds.
If one needs to know the toric data
for the divisor 
which can genarate the superpotential,
one has only to use the techniques
obtained in \cite{sa3,sa4}.
We will consider the illustrative examples
in the next section.

\section{Applications}
As a first example, we consider the \lg\ model
which corresponds to
the hypersurface embedded in $WCP_{(1,1,1,1,4,4)}[12]$
\begin{equation}
W_{1} = 
X_{1}^{12} +
X_{2}^{12} +
X_{3}^{12} +
X_{4}^{12} +
X_{5}^{3} +
X_{6}^{3} .
\end{equation}
This model is already studied in \cite{witten,mayr}.
By our \lg\ analysis, we first find two $(-1,1)$ states
$\keta {j^{-1}}$
and
$\keta {j^{-3}}$.
The state $\keta {j^{-1}}$ 
corresponds to the canonical divisor.
We should pay attention to the fact
that only two fields $X_{5}$ and $X_{6}$
are invariant under $j^{-3}$ action.
So the state $\keta {j^{-3}}$ correponds to 
the Case A in the previous section
and the index of (\ref{beta}) is obtained to be
$\beta(j^{-3}) = 2$
(note that the index $\beta$ is calculated
over the $(c,c)$ ring).
This implies there are more two $(-1,1)$ states,
namely
${X_{5}}{\keta {j^{-2}}}$ and 
${X_{6}}{\keta {j^{-2}}}$.
For divisors this implies
$h^{0,0}({\cal D}) = 3$,
i.e. three independent divisors intersect 
the \cy\ hypersurface in the same way \cite{mayr}.

To resolve this unsatisfactory situation,
we apply the orbifold construction developped in \cite{sa4}.
We orbifoldize this model by the $\ZZ_{3}$ twist
$g = {\rho_{5}^{2}}{\rho_{6}}$,
where ${\rho_{i}}{X_{j}} = 
{e^{2 \pi i {q_{i}} {\delta_{i,j}} }}{X_{j}}$.
The orbifoldized potential is obtained to be
\begin{equation}
W_{1}^{\prime} = 
{X^{\prime}}_{1}^{12} +{X^{\prime}}_{2}^{12}+
{X^{\prime}}_{3}^{12} +{X^{\prime}}_{4}^{12} +
{{X^{\prime}}_{5}^{2}} {{X^{\prime}}_{6}} +
{{X^{\prime}}_{5}} {{X^{\prime}}_{6}^{2}}.
\end{equation}
The two $(-1,1)$ states
${X_{5}}{\keta {j^{-2}}}$ and 
${X_{6}}{\keta {j^{-2}}}$ 
in the original theory
are mapped to in the orbifoldized theory
$\keta {{g^{-1}}{j^{-2}}}$ and
$\keta {{g^{-2}}{j^{-2}}}$,
respectively.
Applying the method in \cite{sa4},
the toric data for the $(-1,1)$ state
${\keta {j^{-3}}}$,
$\keta {{g^{-1}}{j^{-2}}}$ and
$\keta {{g^{-2}}{j^{-2}}}$
are obtained to be 
$(0,0,0,-1,-1)$,
$(0,0,0,1,0)$ and
$(0,0,0,0,1)$,
respectively.
Thus the three divisors,
which can not have the toric data independently
in the original theory,
are represented independently
in the orbifoldized theory.
All of these divisors with wrapped 5-branes
can generate superpotentials.
These two theories of $W_{1}$ and ${W^{\prime}}_{1}$
are conected by this orbifold transition
as was shown in \cite{sa4}.

Our next example is the \lg\ model whose potential is
\begin{equation}
W_{2} = 
X_{1}^{24} +
X_{2}^{24} +
X_{3}^{12} +
X_{4}^{6} +
X_{5}^{3} +
X_{6}^{3},
\end{equation}
which corresponds to the hypersurface
embedded in $WCP_{(1,1,2,4,8,8)}[24]$.
This is the three-fold fibered \cy\ four-fold
\cite{bs}
and its Hodge numbers are
$h^{1,1} = 6$,
$h^{3,1} = 803$,
$h^{2,1} = 1$ and 
$h^{2,2} = 3,278$.

We will concentrate on the pair of states
$\keta {j^{-6}}$ of $U(1)$ charge $(-1,1)$ and
${X_{4}}{\keta {j^{-5}}}$ of $(-2,1)$,
since this pair corresponds to the Case B
in the previous section.
Thus the divisor corresponding to 
the $(-1,1)$ state $\keta {j^{-6}}$
cannot generate a superpotential.
The index of (\ref{beta2}) is obtained to be
$\beta(j^{-6}) = 1$,
as expected.
The toric data of 
the $(-1,1)$ state
$\keta {j^{-6}}$ is obtained to be
$(0,0,-1,-2,-2)$.

We will show that 
after taking an appropriate orbifoldization
this divisor can generate a superpotential.
If we apply the appropriate orbifoldization,
then the $(-2,1)$ state
${X_{4}}{\keta {j^{-5}}}$
is projected out,
so that the divisor corresponds to 
the $(-1,1)$ state
$\keta {j^{-6}}$
{\it can} generate the superpotential.
We can easily find such an appropriate orbifoldization
by the method obtained in \cite{sa4}.

In this case,
we should consider the $\ZZ_{3}$ orbifoldization 
by the twist
$g = {{\rho_{4}^{2}}{\rho_{5}}{\rho_{6}}}$.
In the orbifoldized theory,
the $(-2,1)$ state
${X_{4}}{\keta {j^{-5}}}$
is projected out
and
the index $\beta(j^{-6})$ is calculated to be
$\beta(j^{-6}) = 0$
as expected.
Thus the divisor corresponds to 
the $(-1,1)$ state
$\keta {j^{-6}}$
{\it can} generate a superpotential.
As was shown in \cite{sa4},
the original and the orbifoldized theories
are connected.
This phase transition can be characterized
by the property of the divisors
which lead to 
the non-perturbatively generated superpotentials.

As usual, new twisted states will appear
in the orbifoldized theory.
For our model,
two new $(-2,1)$ states
$\keta {{g^{-1}}{j^{-23}}}$ and
$\keta {{g^{-2}}{j^{-23}}}$
arise.
So the new divisors which correspond to
the new $(-1,1)$ states
$\keta {{g^{-1}}}$ and
$\keta {{g^{-2}}}$
cannot generate the superpotentials.

It is interesting to note that
the $(-2,1)$ state
${X_{4}}{\keta {j^{-5}}}$
in the original theory
is mapped to 
the $(-1,1)$ state
${\keta {{g^{-1}}{j^{-5}}}}$
which can generate the superpotential.
In general,
by the $U(1)$ charge analysis of \cite{sa4},
we see that the $(-p,1)$ states
in the original theory
written in the form
$\prod_{\thd = 0}{X_{i}^{l_{i}}} \keta h$
for $p = 1, \ 2, \ 3$
can be mapped to the $(-1,1)$ states
$\keta {{g^{-l}}{h}}$
in our orbifoldized theory.
In other words,
if there exists
one divisor in the original four-fold
which cannot generate the superpotential,
we obtain several divisors 
in the orbifoldized four-fold, 
at least one of which
can generate a superpotential.

\vspace{1cm}

{\it Acknowledgements} : 
The author would like to thank H. Itoyama
for valuable discussions
and his continuous encouragement.


\end{document}